\documentclass[nohyper,12pt,letterpaper]{JHEP3}
\usepackage{epsfig}
\usepackage[latin1]{inputenc}
\usepackage{bbm,amsfonts}

\author{M. Pirrone\\
Dipartimento di Fisica, Universit\`a di Milano--Bicocca and INFN, Sezione di Milano--Bicocca,
Piazza della Scienza 3, I-20126 Milano, Italy\\
\qquad\\
E-mail: \email{marco.pirrone@mib.infn.it}}

\abstract{We study giant graviton probes in the framework of the three--parameter deformation of the $AdS_5\times S^5$ background. We examine both the case when the brane expands in the deformed $\tilde{S}^5$ part of the geometry and the case when it blows up into $AdS_5$. Performing a detailed analysis of small fluctuations around the giants, the configurations turn out to be stable. Our results hold even for the supersymmetric Lunin--Maldacena deformation.}

\preprint{Bicocca-FT-06-8}

\title{Giants On Deformed Backgrounds}


\csname @addtoreset\endcsname{equation}{section}

\def\bseq{\begin{subequation}}  % = 1a 1b
\def\eseq{\end{subequation}}
\def\bsea{\begin{subeqnarray}}  % = 1.1a 1.1b
\def\esea{\end{subeqnarray}}
\newcommand{\beq}{\begin{equation}}
\newcommand{\bea}{\begin{eqnarray}}
\newcommand{\eea}{\end{eqnarray}}
\newcommand{\eeq}{\end{equation}}

\newcommand {\non}{\nonumber}

\newcommand{\mc}{\mathcal}
\newcommand{\vs}[1]{\vspace{#1 mm}}

\renewcommand{\b}{\beta}

\def\Mb{\kern 2pt\mathchoice
        {%displaystyle
         \vbox{\hrule width10pt height 0.4pt depth 0pt
         \kern 1.2pt\hbox{\kern -2pt$\displaystyle M$}}}
        {%textstyle
         \vbox{\hrule width10pt height 0.4pt depth 0pt
         \kern 1.2pt\hbox{\kern -2pt$\textstyle M$}}}
        {%scriptstyle \kern 0.5pt
\vbox{\hrule width6pt height 0.4pt depth 0pt
         \kern 1.0pt\hbox{\kern -2pt$\scriptstyle M$}}}
        {%scriptscriptstyle \kern 0.5pt
         \vbox{\hrule width5pt height 0.4pt depth 0pt
         \kern 0.8pt\hbox{\kern -2pt$\scriptscriptstyle M$}}}}

\def\Sb{\kern 2pt\mathchoice
        {%displaystyle
         \vbox{\hrule width6pt height 0.4pt depth 0pt
         \kern 1.2pt\hbox{\kern -2pt$\displaystyle S$}}}
        {%textstyle
         \vbox{\hrule width6pt height 0.4pt depth 0pt
         \kern 1.2pt\hbox{\kern -2pt$\textstyle S$}}}
        {%scriptstyle
         \vbox{\hrule width3.5pt height 0.4pt depth 0pt
         \kern 1.0pt\hbox{\kern -2pt$\scriptstyle S$}}}
        {%scriptscriptstyle
         \vbox{\hrule width3pt height 0.4pt depth 0pt
         \kern 0.8pt\hbox{\kern -2pt$\scriptscriptstyle S$}}}}

\def\Rb{\kern 2pt\mathchoice
        {%displaystyle
         \vbox{\hrule width5.5pt height 0.4pt depth 0pt
         \kern 1.2pt\hbox{\kern -2.5pt$\displaystyle R$}}}
        {%textstyle
         \vbox{\hrule width5.5pt height 0.4pt depth 0pt
         \kern 1.2pt\hbox{\kern -2.5pt$\textstyle R$}}}
        {%scriptstyle
         \vbox{\hrule width3.5pt height 0.4pt depth 0pt
         \kern 1.0pt\hbox{\kern -2.2pt$\scriptstyle R$}}}
        {%scriptscriptstyle
         \vbox{\hrule width3pt height 0.4pt depth 0pt
         \kern 0.8pt\hbox{\kern -2.2pt$\scriptscriptstyle R$}}}}

  \def\pp{{\mathchoice
      {%displaystyle
          \kern 1pt%
          \raise 1pt
          \vbox{\hrule width5pt height0.4pt depth0pt
            \kern -2pt
            \hbox{\kern 2.3pt
              \vrule width0.4pt height6pt depth0pt
              }
            \kern -2pt
            \hrule width5pt height0.4pt depth0pt}%
            \kern 1pt
       }
        {%textstyle
          \kern 1pt%
          \raise 1pt
          \vbox{\hrule width4.3pt height0.4pt depth0pt
            \kern -1.8pt
            \hbox{\kern 1.95pt
              \vrule width0.4pt height5.4pt depth0pt
              }
            \kern -1.8pt
            \hrule width4.3pt height0.4pt depth0pt}%
            \kern 1pt
        }
        {%scriptstyle
          \kern 0.5pt%
          \raise 1pt
          \vbox{\hrule width4.0pt height0.3pt depth0pt
            \kern -1.9pt  %[e]=0.15pt
            \hbox{\kern 1.85pt
              \vrule width0.3pt height5.7pt depth0pt
              }
            \kern -1.9pt
            \hrule width4.0pt height0.3pt depth0pt}%
            \kern 0.5pt
        }
        {%scriptscriptstyle
          \kern 0.5pt%
          \raise 1pt
          \vbox{\hrule width3.6pt height0.3pt depth0pt
            \kern -1.5pt
            \hbox{\kern 1.65pt
              \vrule width0.3pt height4.5pt depth0pt
              }
            \kern -1.5pt
            \hrule width3.6pt height0.3pt depth0pt}%
            \kern 0.5pt%}
        }
    }}

  \def\mm{{\mathchoice
               {%displaystyle
                 \kern 1pt
           \raise 1pt    \vbox{\hrule width5pt height0.4pt depth0pt
                  \kern 2pt
                  \hrule width5pt height0.4pt depth0pt}
                 \kern 1pt}
               {%textstyle
                \kern 1pt
           \raise 1pt \vbox{\hrule width4.3pt height0.4pt depth0pt
                  \kern 1.8pt
                  \hrule width4.3pt height0.4pt depth0pt}
                 \kern 1pt}
               {%scriptstyle
                \kern 0.5pt
           \raise 1pt
                \vbox{\hrule width4.0pt height0.3pt depth0pt
                  \kern 1.9pt
                  \hrule width4.0pt height0.3pt depth0pt}
                \kern 1pt}
               {%scriptscriptstyle
               \kern 0.5pt
         \raise 1pt  \vbox{\hrule width3.6pt height0.3pt depth0pt
                  \kern 1.5pt
                  \hrule width3.6pt height0.3pt depth0pt}
               \kern 0.5pt}
               }}

\def\pd{{\kern0.5pt
           + \kern-5.05pt \raise5.8pt\hbox{$\textstyle.$}\kern
0.5pt}}

\def\pmd{{\kern0.5pt
          \pm \kern-5.05pt
\raise6.3pt\hbox{$\textstyle.$}\kern1.5pt}}

\def\md{{\mathchoice
   {%displaystyle
      {{\kern 1pt - \kern-6.2pt \raise5pt\hbox{$\textstyle.$}\kern
1pt}}}
    {%textstyle
      {{\kern 1pt - \kern-6.2pt \raise5pt\hbox{$\textstyle.$}\kern
1pt}}}
    {%scriptstyle
      {\kern0.5pt - \kern-5.05pt
\raise3.4pt\hbox{$\textstyle.$}\kern0.5pt}}
    {%scriptscriptstyle
      {\kern0.5pt - \kern-5.05pt
\raise3.4pt\hbox{$\textstyle.$}\kern0.5pt}}}}

\begin{document}

\section{Introduction}

The standard $AdS/CFT$ correspondence \cite{AdS/CFT} relates a four--dimensional $\mathcal{N}=4$ $SU(N)$ superconformal gauge theory to a Type IIB string theory on $AdS_5\times S^5$. The string theory side of the duality could be studied in its low energy effective description, i.e. in terms of Type IIB classical supergravity. The regime of validity of this approximation requires a small curvature of the background compared to the string scale which is well known to be incompatible to the perturbative regime of the dual gauge theory. So this strong/weak duality is very difficult to test. However a large number of checks have been successfully performed and the impressive amount of evidence supporting it suggests that we have a powerful tool to understand the strongly coupled sector of a gauge theory. The correspondence can be extended to more realistic theories characterized by less supersymmetries with respect to the original formulation and even in situations with non--conformal symmetry, to the aim of studying QCD--like theories. At the moment we are far from a quantitatively understanding of non--perturbative aspects of QCD\footnote{It is a general expectation that the dual of pure QCD could be a strongly coupled string model.}. However it is possible to use the correspondence to learn about the properties of field theories which were previously only poorly understood.   

We are interested in theories with less (or no) supersymmetries which preserve their conformal feature. Starting from an $\mathcal{N}=4$ SYM, if $\mc N=1$ superconformal invariance is required the field theory can be realized by the exactly marginal deformations of the $\mc N=4$ SYM first classified in \cite{LS}. In \cite{LM} Lunin and Maldacena found the gravity dual of the so called $\b$--deformed theory. In the case of real deformation parameter $\beta\equiv\gamma$ the new $AdS_5\times \tilde{S}^5$ background can be obtained from the original  $AdS_5\times S^5$ solution by applying a TsT (T--duality, shift, T--duality) transformation in $S^5$. A natural non--supersymmetric generalization of the Lunin--Maldacena background has been obtained in \cite{F} by performing a series of TsT transformations on each of the three tori of $S^5$ but with different shift parameters $\hat{\gamma}_i$. This background is believed to be dual to a non--supersymmetric but still conformal gauge theory obtained by a related three--parameter deformation of the $\mathcal{N}=4$ SYM. If all the $\hat{\gamma}_i$ are equal, the deformation reduces to the Lunin--Maldacena one. Other interesting generalizations can be found in \cite{RVY,AAF}. A considerable effort has been devoted so far to provide tests of the $AdS/CFT$ in its marginal deformed version and the general idea is to follow what has been done in the original correspondence. The first check has been obtained in \cite{T}.

The gauge/gravity duality was constructed using D3--branes, so it is clear that stable configurations of D3--branes play an important role in this context. Inspired by the work of Myers \cite{My}, the authors of \cite{MGST} found an expanded brane configuration in the $AdS_5\times S^5$ background with exactly the same quantum numbers of a point particle: The giant graviton. It was described as a D3--brane sitting at the center of $AdS_5$, wrapping an $S^3$ onto the $S^5$ part of the geometry and traveling around an equator of the internal space. The main feature of the giant graviton is its stability and the relation between its radius and its angular momentum. Since the radius of the giant cannot be greater than the radius of the space--time, there is an upper bound for the momentum of the brane, the so called stringy exclusion principle. In \cite{GMT,HHI} it was shown that also stable configurations blown up into the $AdS$ part of the geometry exist: The dual giant gravitons. In this case, they have a completely different behavior due to the fact that the $AdS$ space--time is non--compact and then there are no constraints on their size. A remarkable fact is that both the configurations saturate a BPS bound for their energy, which turns out to be equal to their angular momentum in units of the radius of the background. The BPS bound follows from their embedding in a supersymmetric theory because they preserve half of the supersymmetries involved \cite{GMT,HHI}. This makes (dual) giant graviton a natural object to study in the framework of $AdS/CFT$ correspondence. A lot is known from the field theory side \cite{BBNS,CJR,others} and the elegant description of these states in terms of free fermions \cite{Bere} has led to a complete classification of all the half--BPS solutions of Type IIB supergravity \cite{LLM}. Other results on giant gravitons can be found in \cite{general}. 

In \cite{KISS} giant graviton configurations were analyzed on the non--supersymmetric three--parameter deformation of the $AdS_5\times S^5$ background. They did not find energetically favorable solutions making the giants unstable states. On the other hand, they showed a striking quantitative agreement between the open string sigma model and the open spin chain arising from the Yang--Mills theory. Moreover, as noted in the recent paper \cite{HM} it seems strange that giant gravitons have been not found in the supersymmetric $\hat{\gamma}_i=\hat{\gamma}$ Lunin--Maldacena background yet and it would be also interesting to study giants which expand in $AdS$ directions. In this article we try to shed light on these problems, revisiting the construction of (dual) giant gravitons in the three--parameter deformed background. Our results can be easily translated to the superconformal Lunin--Maldacena deformation by setting $\hat{\gamma}_i=\hat{\gamma}$.

The plan of the paper is as follows. After an introductory section on the $\hat{\gamma}_i$--deformed background, in Section 3 we propose an analysis from a point--particle point of view to understand how (and if) the deformation manifests itself in the study of geodesics of the deformed background. In Section 4 we give an ansatz for extended brane solutions blown up in the deformed $\tilde{S}^5$ part of the geometry (giant gravitons) and also in the $AdS_5$ space--time (dual giant gravitons). We find potentially stable states in both cases and an identical scenario to the undeformed one where (dual) giant gravitons behave as point--like gravitons. We note that the symmetric $\hat{\gamma}_i=\hat{\gamma}$ case is not special as long as the procedure seems to be independent of the specific value of the deformation parameters. In Sections 5 we prove that our giants are effectively solutions which minimize the action. Moreover, we examine the bosonic spectrum of small fluctuations around the classical solutions where the deformation of the background plays a crucial role and we show that all fluctuation modes have real frequencies. This signals that (dual) giant gravitons are stable over perturbation even in the presence of non--vanishing $\hat{\gamma}_i$ parameters. In Section 6 we compare our Dirac--Born--Infeld results with qualitative and, where possible, quantitative expectations from the dual $CFT$ pictures. The main focus of this section is on possible directions along which our work can be extended. Then we summarize and conclude.

\section{Generalities on the three--parameter deformation of $AdS_5\times S^5$}
 
The Type IIB supergravity background we will study is related by T--dualities and shift transformations to the usual $AdS_5\times S^5$ and is the generalization of the background first proposed in \cite{LM} to the case of three unequal $\hat{\gamma}_i$ parameters \cite{F}. The corresponding background is a non--supersymmetric deformation of $AdS_5\times S^5$ and should be dual to a non--supersymmetric but marginal deformation of $\mathcal{N}=4$ SYM. Since the deformation is exactly marginal, the $AdS$ factor remains unchanged. The metric of the so called $AdS_5\times\tilde{S}^5$ solution (written in string frame and with $\alpha'=1$) can be read from

\beq\label{metric}
ds^2=ds^2_{AdS_5}+ds^2_{\tilde{S}^5}
\eeq

\noindent
where

\beq\label{ads}
ds^2_{AdS_5}=-(1+\frac{l^2}{R^2})dt^2+\frac{dl^2}{1+\frac{l^2}{R^2}}+l^2\left[d\alpha_1^2+\sin^2{\alpha_1}\left(d\alpha_2^2+\sin^2{\alpha_2}d\alpha_3^2\right)\right]
\eeq

\noindent
is the usual $AdS_5$ space--time and 

\beq\label{sdef}
ds^2_{\tilde{S}^5}=R^2\left(\frac{dr^2}{R^2-r^2}+\frac{r^2}{R^2}d\theta^2+G\sum_{i=1}^{3} \rho_i^2 d\varphi_i^2\right) + R^2 G \rho_1^2 \rho_2^2 \rho_3^2 \left(\sum_{i=1}^{3}\hat{\gamma}_i d\varphi_i\right)^2
\eeq

\noindent
is the deformed five--sphere. Here

\beq
G^{-1}=1+\hat{\gamma}_1^2 \rho_2^2 \rho_3^2+\hat{\gamma}_2^2 \rho_1^2 \rho_3^2+\hat{\gamma}_3^2 \rho_1^2 \rho_2^2\,, \qquad \qquad \hat{\gamma}_i=R^2 \gamma_i
\eeq

\noindent
and it is convenient to parametrize $\rho_i$ coordinates via $\rho_1^2=1-\frac{r^2}{R^2}\,,\,\,\,\rho_2^2=\frac{r^2}{R^2}\cos^2{\theta}\,,\,\,\,\rho_3^2=\frac{r^2}{R^2}\sin^2{\theta}$. Note that $\sum_{i=1}^3 \rho_i^2=1$ and we have $0\leq r \leq R$. We consider only the case of real deformation parameters $\hat{\gamma}_i$, when the axion field $\chi$ is a constant and is set to zero. With respect to the dilaton $\phi_0$ of the undeformed background, the dilaton  $\phi$ of the solution is

\beq
e^{2\phi}=e^{2\phi_0}G
\eeq

\noindent
and we have the usual $AdS/CFT$ relation $R^4=4\pi e^{\phi_0}N=\lambda$, relating the radius of the background and the 't Hooft coupling constant. Note that the dilaton field $\phi$ is not simply a constant, but it depends on the coordinates of the deformed sphere $\tilde{S}^5$.

There is a non--zero NS--NS two form

\beq\label{B}
B=R^2 G \left(\hat{\gamma}_3 \rho_1^2 \rho_2^2 d\varphi_1 \wedge d\varphi_2+\hat{\gamma}_1\rho_2^2\rho_3^2 d\varphi_2\wedge d\varphi_3+\hat{\gamma}_2\rho_3^2\rho_1^2 d\varphi_3\wedge d\varphi_1   \right)\,,
\eeq

\noindent
while the R--R forms are

\beq
C_2=-4 R^2 e^{-\phi_0} \omega_1 \wedge\sum_{i=1}^{3}\hat{\gamma}_i d\varphi_i\,,\qquad\qquad d\omega_1=\frac{r^3}{R^4} \sin{\theta} \cos{\theta}\, dr\wedge d\theta
\eeq

\noindent
and

\bea
C_4&=& e^{-\phi_0} \frac{l^4}{R}\sin^2{\alpha_1}\sin{\alpha_2}dt\wedge d\alpha_1 \wedge d\alpha_2 \wedge d\alpha_3+\non\\
&+& 4 R^4 e^{-\phi_0} G \,\omega_1\wedge d\varphi_1 \wedge d\varphi_2 \wedge d\varphi_3
\eea

\noindent
The five form field strength of the background is

\beq
F_5=dC_4-C_2\wedge dB\,\qquad\qquad \ast F_5=F_5
\eeq

\noindent
When all the three deformation parameters are equal, $\hat{\gamma}_i=\hat{\gamma}$, we recover the Lunin--Maldacena supersymmetric background \cite{LM}.

\section{A rotating point particle probe}\label{g-p}

As a warm up for what follows, we focus on the motion of a massless point--like particle in the deformed $AdS_5\times \tilde{S}^5$ background which rotates on the $\tilde{S}^5$ and minimizes its energy in this internal space. For convenience we start from the action for a massive particle in ten dimensions and later take the mass $M$ to zero,

\beq
S=-M\int dt \sqrt{-(g-b)}
\eeq

\noindent
where $g$ and $b$ are, respectively, the pull--backs of the space--time metric and of the NS--NS two form onto the particle's worldline and are given by

\beq
g=G_{MN}\dot{X}^M\dot{X}^N\,\qquad\qquad b=B_{MN}\dot{X}^M\dot{X}^N
\eeq

\noindent
Here $X^M$ are coordinates on the ten--dimensional space--time with $X^0=t$ and $\dot{X}^M$ denotes the derivative of $X^M$ with respect to $t$. The metric $G_{MN}$ and the NS--NS two form $B_{MN}$ can be read in (\ref{metric}) and in (\ref{B}), respectively. The rotating point particle we want to analyze sits at the center of $AdS_5$ and spins in the $\varphi_1$ direction. For this configuration we have $g=G_{tt}+G_{\varphi_1\varphi_1}\,\dot{\varphi}_1^2$, $\,b=0$ and the action becomes 

\beq
S=-M\int dt \sqrt{1-R^2 G \rho_1^2(1+\hat{\gamma}_1^2 \rho_2^2 \rho_3^2)\dot{\varphi}_1^2}
\eeq

\noindent
From now on, to save space we introduce the positive quantity $Q^2=R^2 G \rho_1^2(1+\hat{\gamma}_1^2 \rho_2^2 \rho_3^2)$. Since the action we have written down presents no explicit dependence on the cyclic coordinate $\varphi_1$, we can replace $\dot{\varphi}_1$ with its conjugate momentum

\beq
J=\frac{\partial L}{\partial \dot{\varphi}_1}=\frac{Q^2 M \dot{\varphi}_1}{\sqrt{1-Q^2\dot{\varphi}_1^2}}
\eeq

\noindent
which is conserved in time. So we can define the Hamiltonian in the standard way

\beq\label{H}
H=\dot{\varphi}_1 J-L=\frac{J}{Q}
\eeq

\noindent
where we have already taken the limit $M\rightarrow 0$. We need to find the minimum of the Hamiltonian and it is easy to convince that this occurs when $Q$ is maximum, namely when $r=0$ and so $Q=R$. Substituting this value in equation (\ref{H}) we obtain the energy of the rotating point particle
 
\beq
E=\frac{J}{R}
\eeq

\noindent
Finally, we find a geodesic which represents a BPS state\footnote{In all our discussions we use the term BPS in its original sense. We do not refer to supersymmetry.} with energy $E$ equal to the angular momentum $J$ (in units of $1/R$) and does not depend on the deformation parameters, i.e. is the same as in the undeformed theory. This is one of the cases already analyzed in \cite{FRT} (see also \cite{AAF}).

\section{The equilibrium configurations}

Our main purpose is to probe the deformed and non--supersymmetric background with giant gravitons. We want to understand if it is possible to find minimum energy configurations, study their stability and eventually their dependence on the deformation parameters. Recall that in the standard $AdS_5\times S^5$ background there are three different configurations characterized by the same quantum numbers. The first one is a point--like graviton spinning around an $S^1$ direction contained in $S^5$, then there is a giant graviton corresponding to a D3--brane wrapping an $S^3\subset S^5$ and the third one is the so called dual giant graviton with the topology of an $S^3\subset AdS_5$. What about the deformed case? 

In general, the dynamics of a D3--brane in a given background is described by the action

\beq\label{dbiwz}
S=S_{DBI}+S_{WZ}
\eeq

\noindent
where the Dirac--Born--Infeld term is

\beq\label{sdbi}
S_{DBI}=-T_3\int_{\Sigma_4} d\tau d^3\sigma\, e^{-\phi}\sqrt{-det(g_{ab}+\mathcal{F}_{ab})}
\eeq

\noindent
With $g_{ab}=G_{MN}\partial_a X^M \partial_b X^N$ we mean the pull--back of the ten--dimensional space--time metric $G_{MN}$ on the worldvolume $\Sigma_4$ of the brane. $T_3$ is the D3--brane tension\footnote{In our conventions $T_3=\frac{1}{(2\pi)^3}$, see \cite{Sz} for example.}. The gauge potential $A_a$ enters the action through a $U(1)$ worldvolume gauge field strength $F_{ab}$ in the modified field strength $\mathcal{F}_{ab}=2\pi F_{ab}-b_{ab}$, where $b_{ab}$ is the pull-back to the worldvolume of the target NS--NS two-form potential, $b_{ab}=B_{MN}\partial_a X^M \partial_b X^N$. D--branes are charged under R--R potentials and this feature determines that their action should contain a term (the Wess--Zumino term) coupling the brane to these fields,

\beq\label{swz}
S_{WZ}=T_3\int_{\Sigma_4}P\left[\sum_{q}C_q\,e^{-B}\right]\,e^{2\pi F}
\eeq

\noindent
where $P[...]$ denotes again the pull--back and the wedge--product is implicit.

Our analysis focuses on purely scalar solutions, so we drop all the fermions and the gauge potential $A_a$ on the brane, as done in the undeformed case.

\subsection{Branes expanding in the deformed $\tilde{S}^5$ space--time: Giant gravitons}\label{confs}
 
The first solutions we want to study are D3--branes wrapped on the deformed sphere part of the geometry, moving entirely in the $\tilde{S}^5$ and sitting at the center of $AdS_5$. The time coordinate in $AdS_5$ is denoted by $t$. In what follows it is convenient to choose a static gauge such that the worldvolume coordinates of the brane $(\tau,\sigma_i)$ are identified with the appropriate space--time coordinates. In particular the brane wraps the $(\theta,\varphi_2,\varphi_3)$ directions,  

\bea
\tau=t\,,\qquad\sigma_1=\theta\in \left[0,\frac{\pi}{2}\right]\,,\qquad\sigma_2=\varphi_2\in \left[0,2\pi\right]\,,\qquad\sigma_3=\varphi_3\in \left[0,2\pi\right]
\eea

\noindent
The D3--brane action (\ref{dbiwz}) can be rewritten as

\beq
S=-T_3 \int_{\Sigma_4}dt d\theta d\varphi_2 d\varphi_3\, e^{-\phi}\sqrt{-det(g_{ab}-b_{ab})}+T_3\int_{\Sigma_4}P\left[C_4-C_2\wedge B\right]
\eeq

\noindent
Our giant graviton has constant radius ($r_0$), it orbits the $\tilde{S}^5$ in the $\varphi_1$ direction with a constant angular velocity ($\omega_0$) and all the worldvolume modes are frozen. While it is not a priori obvious that this is a consistent way of embedding the brane, we will see that it gives in fact a minimal energy configuration. So we propose an ansatz of the form

\beq
r=r_0\,\qquad \varphi_1=\omega_0 t\,\qquad l=\alpha_1=\alpha_2=\alpha_3=0
\eeq

\noindent
which, after integration on the spatial coordinates of the worldvolume, leads to the effective Lagrangian

\beq
L=-h\sqrt{1-a^2 \dot{\varphi_1}^2}+m\dot{\varphi}_1
\eeq

\noindent
with 

\beq
h= N \frac{r_0^3}{R^4}\,,\qquad a^2=R^2-r_0^2\,,\qquad m= N \frac{r_0^4}{R^4}
\eeq

\noindent
We have the constraint $r_0\leq R$ because the size of the brane cannot exceed the radius of $\tilde{S}^5 $ and so $a^2\geq 0$. We have also used $A_3\, T_3 \,e^{-\phi_0}=\frac{N}{R^4}$, where $A_3$ is the area of a unit 3--sphere. Note that the effective Lagrangian is exactly the same found in the undeformed case \cite{MGST,GMT,HHI} and this appears to be strange at first sight because the giant has blown up in the deformed $\tilde{S}^5$. We will comment later on this particular behavior which is in contrast with the results obtained in \cite{KISS}.

The conjugate momentum to $\varphi_1$ is

\beq
J=\frac{\partial L}{\partial \dot{\varphi}_1}=\frac{h a^2 \dot{\varphi_1}}{\sqrt{1-a^2 \dot{\varphi_1}^2}}+m
\eeq
 
\noindent
This relation can be easily inverted to obtain

\beq
\dot{\varphi_1}=\frac{J-m}{a^2\sqrt{h^2+\frac{(J-m)^2}{a^2}}}
\eeq

\noindent
The corresponding Hamiltonian of the giant graviton becomes

\beq\label{ham}
H=\dot{\varphi_1} J-L=\sqrt{h^2+\frac{(J-m)^2}{a^2}}
\eeq

\noindent
and it is independent of $\varphi_1$, so that the equations of motion can be solved with constant momentum. For fixed $J$, we have two extrema of (\ref{ham}) now regarded as the potential that determines the equilibrium radius. In particular, there are two degenerate minima at $r_0=0$ and at $r_0=R\sqrt{\frac{J}{N}}$, where the energy is $E=\frac{J}{R}$, as for the point graviton, and $\omega_0=\dot{\varphi_1}=\frac{1}{R}$. This analysis obviously gives the same results already found in the undeformed case and the stringy exclusion principle manifests itself in the relation between the radius of the giant and its angular momentum.

\subsection{Branes expanding in $AdS_5$ space--time: Dual giant gravitons}\label{confads}

In the previous section we have seen that there is a D3--brane configuration with the same quantum numbers as the point--like graviton, even in the deformed $AdS_5\times\tilde{S}^5$ background. Now we also consider the possibility of dual giant graviton solutions where the D3--branes are wrapped in the 3--sphere $(\alpha_1,\alpha_2,\alpha_3)$ contained in the $AdS_5$ part of the geometry. In contradistinction to the previous case we expect a priori the effective Lagrangian not to depend on the deformation parameters because they do not enter the $AdS$ space--time \cite{KISS,HM}. Again the dynamics is described by the action (\ref{dbiwz}) and we use the static gauge for the worldvolume coordinates of the brane $(\tau,\sigma_i)$,

\bea
\tau=t\,,\qquad\sigma_1=\alpha_1\in \left[0,\pi\right]\,,\qquad\sigma_2=\alpha_2\in \left[0,\pi\right]\,,\qquad\sigma_3=\alpha_3\in \left[0,2\pi\right]
\eea

\noindent
The giant graviton has constant radius ($l_0$) and again orbits rigidly in the $\varphi_1$ direction on the $\tilde{S}^5$. Our ansatz is

\beq
l=l_0\,\qquad \varphi_1=\omega_0 t\,\qquad r=\varphi_2=\varphi_3=0\,\qquad \theta=\frac{\pi}{4}
\eeq

\noindent
We will see that with the parametrization of the deformed 5--sphere as in (\ref{sdef}), the choice $\theta=\pi/4$ is the most natural one in the study of fluctuations around the giant. The dependence on the deformation parameters of the vibrations turns out to depend on the position of the giant into the internal space. This ansatz yields the effective Lagrangian

\beq
L =-\tilde{h}\sqrt{\tilde{b}^2-R^2\dot{\varphi_1}^2}+\tilde{m}
\eeq

\noindent
with

\beq\label{parads}
\tilde{h}= N \frac{l_0^3}{R^4}\,,\qquad \tilde{b}^2=1+\frac{l_0^2}{R^2}\,,\qquad \tilde{m}= N \frac{l_0^4}{R^5}
\eeq

\noindent
as in the undeformed case \cite{GMT,HHI}. Again we have used $A_3\, T_3\, e^{-\phi_0} =\frac{N}{R^4}$. The conjugate momentum to $\varphi_1$ now becomes

\beq
J=\frac{\partial L}{\partial \dot{\varphi}_1}=\frac{\tilde{h} R^2 \dot{\varphi_1}}{\sqrt{\tilde{b}^2-R^2 \dot{\varphi_1}^2}}
\eeq

\noindent 
and from this relation we obtain

\beq
\dot{\varphi_1}=\frac{J\tilde{b}}{R^2\sqrt{\tilde{h}^2+\frac{J^2}{R^2}}}
\eeq

\noindent
We can calculate the corresponding Hamiltonian of the dual giant graviton and obtain

\beq
H=\dot{\varphi_1} J-L=\tilde{b}\sqrt{\tilde{h}^2+\frac{J^2}{R^2}}-\tilde{m}
\eeq

\noindent
Again  $H$, as a function of $l_0$, has two minima located at $l_0=0$ and $l_0=R\sqrt{\frac{J}{N}}$. The energy at each minima is $E=\frac{J}{R}$ and $\omega_0=\dot{\varphi_1}=\frac{1}{R}$, matching the results of the previous sections. Of course now there is no upper bound on the angular momentum $J$ because $AdS$ space--time is non--compact and the radius $l_0$ of the giant can be greater than $R$ \cite{GMT,HHI}.  

So far we have seen that even for the deformed background $AdS_5\times \tilde{S}^5$, there are three potential configurations to describe a graviton carrying angular momentum $J$: The point--like graviton, the giant graviton of section \ref{confs} consisting of a 3--brane expanded into the deformed 5--sphere, and a dual giant graviton consisting of a spherical 3--brane which expands into the $AdS$ space. This is exactly the same situation known from the standard undeformed $AdS_5\times S^5$ background. Moreover, if we consider the collective motion of both brane configurations, we see that their center of mass travels along a null trajectory in the ten--dimensional space--time once evaluated in $\dot{\varphi_1}=1/R$. We stress that this is the expected result for a massless point--like graviton, but it is also true for the expanded (dual) giant gravitons. So we have really found that giant graviton states which are degenerated with massless particle states exist classically even in a background which in general preserves no supersymmetries. This result is not so strange because it is a feature of a large class of non--supersymmetric backgrounds \cite{PS} and of particular configurations in theories with non zero NS--NS $B$ field \cite{CRK}. 

\section{Stability analysis and vibration modes}\label{vib}

One of the main issues related to giant gravitons is their stability under the perturbation around the equilibrium configurations. In the last two sections we found expanded branes with the same energy of a point graviton and so they should be stable. In order to verify this expectation we will consider the spectrum of small fluctuations around the giants, as first studied in \cite{DJM}. A vibration of the brane can be described by expanding our previous ansatz as follows

\beq
X=X_0+\varepsilon \delta X(t,\sigma_i)
\eeq

\noindent
where $X$ is a generic space--time coordinate, $X_0$ denotes the solution of the unperturbed equilibrium configuration, the fluctuation $\delta X(t,\sigma_i)$ is a function of the worldvolume coordinates $(t,\sigma_i)$ and $\varepsilon$ is a small perturbation parameter. We work in a Lagrangian setup \cite{DJM} and we expand the action of the probe brane in powers of $\varepsilon$ as

\beq
S=\int dt d^3\sigma\{\mathcal{L}_0+\varepsilon \mathcal{L}_1+\varepsilon^2\mathcal{L}_2+\cdots\}
\eeq

\noindent
Obviously $\mathcal{L}_0$ gives a zeroth order Lagrangian density related to that we have found in the previous sections. To state that those solutions really minimize the action we have to focus on the $\mathcal{L}_1$ term. The second order term $\mathcal{L}_2$ is useful to study the stability of the configurations we have found and the bosonic fluctuation spectrum, which we expect to depend on the deformation parameters, as in the analysis of vibrations around other BPS states of this background \cite{FRT}. Perturbative instability will manifest in the spectrum as a tachyonic mode. We closely follow \cite{DJM}. A slightly different method has been proposed in \cite{O}.

\subsection{Giant graviton fluctuations}\label{ggf}

To study the fluctuations around the configurations found in section \ref{confs} it is useful to rewrite the $AdS_5$ part of the metric as suggested in \cite{DJM}

\beq
ds^2_{AdS_5}=-\left(1+\sum_{k=1}^{4}v_k^2\right)dt^2+R^2\left(\delta_{ij}+\frac{v_i v_j}{1+\sum_{k=1}^{4}v_k^2}\right)dv_i dv_j
\eeq

\noindent
Then we change our previous ansatz as

\beq
r=r_0+\varepsilon\,\delta r(t,\sigma_i)\,\qquad \varphi_1=\omega_0 t+\varepsilon\,\delta \varphi_1(t,\sigma_i)\,\qquad v_k=\varepsilon\,\delta v_k(t,\sigma_i)
\eeq

\noindent
with $\sigma_i=(\theta,\varphi_2,\varphi_3)$. Expanding the action to the linear order we get

\bea\label{l1}
\mathcal{L}_1&=&-T_3 \,e^{-\phi_0}\, \sin{\theta}\cos{\theta}\non\\
&r_0^2&\left\{\left[\frac{4 r_0^2 \omega_0^2+3(1-R^2 \omega_0^2)}{\sqrt{1-(R^2-r_0^2)\omega_0^2}}-4 r_0 \omega_0\right]\delta r+\right.\non\\
&-&\left.\left[\frac{(R^2-r_0^2)r_0 \omega_0}{\sqrt{1-(R^2-r_0^2)\omega_0^2}}+r_0^2\right]\frac{\partial \delta \varphi_1}{\partial t}\right\}
\eea

\noindent
The first order Lagrangian density (\ref{l1}) does not contain the deformation parameters and is exactly the same found in the undeformed analysis \cite{DJM}. The term in front of $\frac{\partial \delta \varphi_1}{\partial t}$ is a constant and so it brings no contribution to the variation of the action with fixed boundary values. The coefficient of the term $\delta r$ vanishes if we take

\beq\label{omega0}
\omega_0=\frac{1}{R}
\eeq

\noindent 
This confirms that the giant graviton described in the previous section (the zeroth order solution) is the right solution which really minimizes the action. Now we consider the second order term in $\varepsilon$. With the choice (\ref{omega0}) we get

\bea
\mathcal{L}_2&=& T_3\,e^{-\phi_0}\,r_0^2\, \sin{\theta}\cos{\theta}\non\\
\Bigg\{\Bigg[&-&\frac{R^3}{2(R^2-r_0^2)}\,\frac{\partial^2 \delta r}{\partial t^2}+\frac{R}{2(R^2-r_0^2)}\,\Delta_{S^3}\,\delta r+\non\\
&+&\frac{1}{2 R}\left(\hat{\gamma}_3^2 \frac{\partial^2 \delta r}{\partial \varphi_2^2}+\hat{\gamma}_2^2 \frac{\partial^2 \delta r}{\partial \varphi_3^2}-2 \hat{\gamma}_2 \hat{\gamma}_3 \frac{\partial^2 \delta r}{\partial \varphi_2 \partial \varphi_3} \right)\Bigg]\delta r+\non\\
\Bigg[&-&\frac{R^3(R^2-r_0^2)}{2 r_0^2}\,\frac{\partial^2 \delta \varphi_1}{\partial t^2}+\frac{R(R^2-r_0^2)}{2 r_0^2}\,\Delta_{S^3}\,\delta \varphi_1\Bigg]\,\delta \varphi_1+\non\\
&+&\frac{2 R^2}{r_0}\frac{\partial \delta \varphi_1}{\partial t}\,\delta r+\non\\
\Bigg[&-&\frac{R^3}{2}\frac{\partial^2 \delta v_k}{\partial t^2}+\frac{R}{2}\,\Delta_{S^3}\,\delta v_k-\frac{R}{2}\,\delta v_k\non\\
&+&\frac{R^2-r_0^2}{2 R}\left(\hat{\gamma}_3^2 \frac{\partial^2 \delta v_k}{\partial \varphi_2^2}+\hat{\gamma}_2^2 \frac{\partial^2 \delta v_k}{\partial \varphi_3^2}-2 \hat{\gamma}_2 \hat{\gamma}_3 \frac{\partial^2 \delta v_k}{\partial \varphi_2 \partial \varphi_3} \right)\Bigg]\delta v_k\Bigg\}
\eea

\noindent
where the sum over $k$ is understood and $\Delta_{S^3}$ is the Laplacian on the unit 3--sphere. In writing $\mathcal{L}_2$ some terms are integrated by parts; there are no surface contributions because the worldvolume of the brane is a closed surface and the variations are assumed to vanish at $t=\pm\infty$. 

Because of the $U(1)\times U(1)$ worldvolume symmetry, corresponding to translations of $\varphi_2$ and $\varphi_3$, it is convenient to introduce spherical harmonics $\mathcal{Y}_s^{m_2,m_3}(\theta,\varphi_2,\varphi_3)$ with definite $U(1)\times U(1)$ quantum numbers $(m_2,m_3)$ \cite{KISS,HM}. In particular we have

\bea
\Delta_{S^3}\,\mathcal{Y}_s^{m_2,m_3}(\theta,\varphi_2,\varphi_3)&=&-Q_s^2 \mathcal{Y}_s^{m_2,m_3}(\theta,\varphi_2,\varphi_3)\non\\
\frac{\partial}{\partial \varphi_{2,3}}\,\mathcal{Y}_s^{m_2,m_3}(\theta,\varphi_2,\varphi_3)&=&i m_{2,3}\mathcal{Y}_s^{m_2,m_3}(\theta,\varphi_2,\varphi_3)
\eea

\noindent
For spherical harmonics on $S^3$, $Q_s^2=s(s+2)$. We expand the perturbations as

\bea
\delta r(t,\theta,\varphi_2,\varphi_3)&=&A_r \,e^{-i \omega t}\,\mathcal{Y}_s^{m_2,m_3}(\theta,\varphi_2,\varphi_3)\non\\
\delta \varphi_1(t,\theta,\varphi_2,\varphi_3)&=&A_{\varphi_1} \,e^{-i \omega t}\,\mathcal{Y}_s^{m_2,m_3}(\theta,\varphi_2,\varphi_3)\\
\delta v_k(t,\theta,\varphi_2,\varphi_3)&=&A_{v_k} \,e^{-i \omega t}\,\mathcal{Y}_s^{m_2,m_3}(\theta,\varphi_2,\varphi_3)\non
\eea

\noindent
The form of $\mathcal{L}_2$ tells us that the $\delta v_k$ perturbations decouple from $\delta r,\,\delta \varphi_1$ and have frequencies given by

\beq\label{omegak}
\omega_k^2=\frac{1}{R^2}\left(1+Q_s^2+\hat{\Gamma}^2\right) 
\eeq

\noindent
where we have defined the positive quantity 

\beq\label{gammaq}
\hat{\Gamma}^2=\left(1-\frac{r_0^2}{R^2}\right)(\hat{\gamma}_3 m_2-\hat{\gamma}_2 m_3)^2
\eeq

\noindent
which contains the whole dependence on the deformation parameters and on the radius $r_0=R\sqrt{J/N}$ of the giant. The fluctuations $\delta r,\,\delta \varphi_1$ are coupled and the resulting frequencies are obtained solving the following matrix equation 

\beq\label{sys}
\left[
\begin{array}{cc}
\frac{R}{R^2-r_0^2}\left(\omega ^2 R^2-Q_s^2-\hat{\Gamma}^2 \right) & -2 i \omega\frac{R^2}{r_0} \\
2 i \omega\frac{R^2}{r_0} & \frac{R(R^2-r_0^2)}{r_0^2}\left(\omega ^2 R^2-Q_s^2\right)
\end{array}
\right]
\left[
\begin{array}{cc} 
A_r \vs{.1} \\ 
A_{\varphi_1} \vs{.1}
 \end{array} 
\right]=0
\eeq

\noindent
The determinant brings us to a quadratic equation for $\omega^2$ from which we obtain

\beq\label{omegapm}
\omega_{\pm}^2=\frac{1}{R^2}\left[2+Q_s^2+\frac{\hat{\Gamma}^2}{2}\pm 2\sqrt{1+Q_s^2+\frac{\hat{\Gamma}^2}{2}\left(1+\frac{\hat{\Gamma}^2}{8}\right)}\,\right]
\eeq

\noindent
The condition for a giant graviton to be stable over the perturbations is that all the frequencies are real, i.e. $\omega^2\geq 0$. The existence of imaginary part in $\omega$ means that the $e^{-i \omega t}$ term can grow exponentially, which gives instability to the configuration, a tachyonic mode. We have the constraint $r_0\leq R$ and so it is easy to conclude that there are not unstable modes in the system at this quadratic order, as all the $\omega^2$ we found are real and nonnegative. Note that because of the deformation parameters these frequencies depend on the radius $r_0$ of the giant (\ref{gammaq}). In the undeformed background all of the frequencies are independent of $r_0$ \cite{DJM} and this is the main difference with respect to the deformed theory.

\subsection{Dual giant graviton fluctuations}\label{dggf}

Now we want to study the fluctuations around the configurations found in section \ref{confads}. The $AdS$ space--time is now better described by the global coordinate metric (\ref{ads}). Hence the ansatz becomes

\beq
l=l_0+\varepsilon\,\delta l(t,\sigma_i)\,\qquad \varphi_1=\omega_0 t+\varepsilon\,\delta \varphi_1(t,\sigma_i)
\eeq

\noindent
and

\beq\label{ans}
r=\varepsilon\,\delta r(t,\sigma_i)\,\qquad \theta=\frac{\pi}{4}+\varepsilon\,\delta \theta(t,\sigma_i)\,\qquad \varphi_2=\varepsilon\,\delta \varphi_2(t,\sigma_i)\,\qquad \varphi_3=\varepsilon\,\delta \varphi_3(t,\sigma_i)
\eeq

\noindent
with $\sigma_i=(\alpha_1,\alpha_2,\alpha_3)$. Expanding the action to the linear order we get the same contribution as in the undeformed background \cite{DJM}

\bea
\mathcal{L}_1&=&- \frac{T_3}{R} \,e^{-\phi_0}\, \sin^2{\alpha_1}\sin{\alpha_2}\non\\
&l_0^2&\left\{\left[\frac{4 l_0^2 +3 R^2(1-R^2 \omega_0^2)}{\sqrt{l_0^2+R^2(1-R^2 \omega_0^2)}}-4 l_0 \right]\delta l+\right.\non\\
&-&\left.\frac{l_0 \omega_0 R^4}{\sqrt{l_0^2+R^2(1-R^2 \omega_0^2)}}\frac{\partial \delta \varphi_1}{\partial t}\right\}\non
\eea

\noindent
Again, the coefficient of the term $\frac{\partial \delta \varphi_1}{\partial t}$ is a constant and so it brings no contribution to the variation of the action with fixed boundary values. The coefficient of the term $\delta l$ vanishes if we take

\beq\label{omega0a}
\omega_0=\frac{1}{R}
\eeq

\noindent
This fact confirms that the giant graviton written in the previous section is a solution to the equation of motion following from the D3--brane action. With this choice the term linear in $\varepsilon$ vanishes, while the second order term is

\bea
\mathcal{L}_2&=& T_3 \,e^{-\phi_0}\,l_0^2\, \sin^2{\alpha_1}\sin{\alpha_2}\non\\
\Bigg\{\Bigg[&-&\frac{R^3}{2(l_0^2+R^2)}\,\frac{\partial^2 \delta l}{\partial t^2}+\frac{R}{2(l_0^2+R^2)}\,\Delta_{S^3}\, \delta l\Bigg]\,\delta l+\non\\
\Bigg[&-&\frac{R^3(l_0^2+R^2)}{2 l_0^2}\frac{\partial^2 \delta \varphi_1}{\partial t^2}+\frac{R(l_0^2+R^2)}{2 l_0^2}\,\Delta_{S^3}\, \delta \varphi_1\Bigg]\, \delta \varphi_1+\non\\
&+&\frac{2 R^2}{l_0}\frac{\partial \delta \varphi_1}{\partial t}\,\delta l+\non\\
\Bigg[&-&\frac{R}{2}\frac{\partial^2 \delta r}{\partial t^2}+\frac{1}{2 R}\,\Delta_{S^3}\, \delta r-\frac{1}{2 R}\left(1+\frac{\tilde{b}^2}{2}\,(\hat{\gamma}_2^2+\hat{\gamma}_3^2)\right)\delta r \Bigg]\,\delta r \Bigg\}
\eea

\noindent
Of course $\Delta_{S^3}$ is the Laplacian on a 3--sphere and $\tilde{b}^2=1+\frac{l_0^2}{R^2}$, as in (\ref{parads}). Let $\tilde{\mathcal{Y}}_s(\alpha_1,\alpha_2,\alpha_3)$ be spherical harmonics so that the usual relation holds

\beq
\Delta_{S^3}\,\tilde{\mathcal{Y}}_s(\alpha_1,\alpha_2,\alpha_3)=-Q_s^2 \tilde{\mathcal{Y}}_s(\alpha_1,\alpha_2,\alpha_3)
\eeq

\noindent
We expand the perturbations as

\bea
\delta l(t,\alpha_1,\alpha_2,\alpha_3)&=&\tilde{A}_l \,e^{-i \tilde{\omega} t}\,\tilde{\mathcal{Y}}_s(\alpha_1,\alpha_2,\alpha_3)\non\\
\delta \varphi_1(t,\alpha_1,\alpha_2,\alpha_3)&=&\tilde{A}_{\varphi_1} \,e^{-i \tilde{\omega} t}\,\tilde{\mathcal{Y}}_s(\alpha_1,\alpha_2,\alpha_3)\\
\delta r(t,\alpha_1,\alpha_2,\alpha_3)&=&\tilde{A}_r \,e^{-i \tilde{\omega} t}\,\tilde{\mathcal{Y}}_s(\alpha_1,\alpha_2,\alpha_3)\non
\eea

\noindent
The $\delta r$ perturbation decouples from $\delta l,\,\delta \varphi_1$ and it has a frequency given by

\beq\label{omegar}
\tilde{\omega}^2_r=\frac{1}{R^2}\left[1+Q_s^2+\frac{\tilde{b}^2}{2}\,(\hat{\gamma}_2^2+\hat{\gamma}_3^2)\right] 
\eeq

\noindent
The $\delta l,\,\delta \varphi_1$ fluctuations are coupled and the resulting normal frequencies are obtained solving

\beq
\left[
\begin{array}{cc}
\frac{R}{l_0^2+R^2}\left(\tilde{\omega}^2 R^2-Q_s^2\right) & -2 i \omega\frac{R^2}{l_0} \\
2 i \omega\frac{R^2}{l_0} & \frac{R(l_0^2+R^2)}{l_0^2}\left(\tilde{\omega}^2 R^2-Q_s^2\right)
\end{array}
\right]
\left[
\begin{array}{cc} 
\tilde{A}_l \vs{.1} \\ 
\tilde{A}_{\varphi_1} \vs{.1}
 \end{array} 
\right]=0
\eeq

\noindent
which yields

\bea\label{omegapma}
\tilde{\omega}_{\pm}^2&=&\frac{1}{R^2}\left(2+Q_s^2\pm 2\sqrt{1+Q_s^2}\right)
\eea

\noindent
Again there are not unstable modes in the system at this quadratic order, as all the frequencies are real. The deformation parameters $\hat{\gamma}_{2,3}$ enter the frequency $\tilde{\omega}^2_r$ which brings a dependence on the radius $l_0=R\sqrt{J/N}$. The frequencies $\tilde{\omega}^2_{\pm}$ are the same as in the undeformed case and do not depend on $l_0$ \cite{DJM}.

\subsection{Summary of the excitation spectrum and role of deformation}

In this section we discuss how the deformation enters the vibration modes. First of all, we stress that turning off $\hat{\gamma}_i$ manifestly reduces all the frequencies to those of the undeformed case. This is a good test of our results.

\begin{itemize}

\item When the giant graviton expands into the deformed sphere, it has six transverse scalar fluctuations, of which four correspond to fluctuations into $AdS_5$ ($\omega^2_k$) and two are fluctuations within $\tilde{S}^5$ ($\omega^2_{\pm}$). In particular from (\ref{omegak}) and (\ref{omegapm})

\bea\label{omega}
\delta v_k \, &\rightarrow& \, \omega^2_k=\frac{1}{R^2}\left(1+Q_s^2+\hat{\Gamma}^2\right)\non\\
(\delta r, \delta \varphi_1) \, &\rightarrow& \, \omega^2_{\pm}=\frac{1}{R^2}\left[2+Q_s^2+\frac{\hat{\Gamma}^2}{2}\pm 2\sqrt{1+Q_s^2+\frac{\hat{\Gamma}^2}{2}\left(1+\frac{\hat{\Gamma}^2}{8}\right)}\,\right]
\eea

\noindent
All the vibrations involve the deformation parameters $\hat{\gamma}_{2,3}$ (\ref{gammaq}) because the perturbations $\delta X(t,\theta,\varphi_2,\varphi_3)$ are functions of the worldvolume coordinates of the brane and in particular they depend on $\varphi_2\,,\varphi_3$. So, once we perturb the giant around the equilibrium configuration in $X_0$ the fluctuations feel the effect of the deformed background. Note that a similar $\hat{\gamma}_{2,3}$ dependence appears also in \cite{FRT} in the calculation of quadratic fluctuations near a $(J,0,0)$ geodesic. The frequencies just discussed are very similar to the ones obtained in \cite{HM}; the main difference is our dependence on the radius of the giant.

\item Similarly, the vibration mode frequencies corresponding to the giant graviton expanded in the $AdS$ part, are (\ref{omegar}) and (\ref{omegapma})

\bea\label{omegatil}
\delta r\,&\rightarrow&\, \tilde{\omega}^2_r=\frac{1}{R^2}\left[1+Q_s^2+\frac{\tilde{b}^2}{2}(\hat{\gamma}_2^2+\hat{\gamma}_3^2)\right]\non\\
(\delta l, \delta \varphi_1) \, &\rightarrow& \, \tilde{\omega}^2_{\pm}=\frac{1}{R^2}\left(2+Q_s^2\pm 2\sqrt{1+Q_s^2}\right)
\eea

\noindent
An accurate analysis of the quadratic expansion tells us that $G_{\varphi_1 \varphi_1}$ brings the whole dependence on the deformation, once one is calculating the pull--back. In section \ref{confads}, we have mentioned that the choice of the parametrization of the $\rho_i$ in (\ref{sdef}) is important in the study of the dual giant vibrations. Physically, their dependence on the deformation is expected due to the location of the giant into the deformed sphere. The coordinates $\rho_i$ are functions of the angle $\theta$ and we are now expanding around $\pi/4$. So, up to $\varepsilon^2$ we obtain $G_{\varphi_1 \varphi_1}\sim R^2-\varepsilon^2(2+\hat{\gamma}_2^2+\hat{\gamma}_3^2)\delta r^2/2$ and the $\hat{\gamma}_{2,3}$ dependence manifests itself only when we study perturbations in $\tilde{S}^5$, as for $\tilde{\omega}^2_r$. The original ansatz $\theta=\pi/4$ does not select a particular deformation parameter. The frequency $\tilde{\omega}^2_r$ is symmetric in the exchange $\hat{\gamma}_2\leftrightarrow\hat{\gamma}_3$ and depends on the radius $l_0$ of the dual giant. On the other hand, we expect independence from the deformation when studying perturbations in $AdS$ directions.
\end{itemize}

\noindent
From (\ref{omega}) and (\ref{omegatil}) we see that $\omega^2_{-}=0$ and $\tilde{\omega}^2_{-}=0$ are solutions when $Q^2_s=0$. These zero modes correspond to the fact that we have no constraints on $r_0$ and $l_0$, namely they can be taken to have any value allowed by the geometry.

\section{Undeformed giants in a deformed background}

At a classical level we have found that the effective Lagrangian and hence the energy of a giant in the $\hat{\gamma}_i$--deformed background are independent of the deformation parameters. This is an expected result for the dual giant (brane expanded in the $AdS$ part of the geometry), but seems quite strange if the brane expands into the deformed 5--sphere. Analytically, this is due to the particular form of the D3--brane action. The kinetic part (\ref{sdbi}) is independent of the deformation because of the presence of the modified dilaton (the same behavior found in \cite{KISS}). The Wess--Zumino part of the action is

\beq
S_{WZ}=T_3 \int_{\Sigma_4} P\left[C_4-C_2 \wedge B\right]
\eeq

\noindent
It is important to note that, even before taking the pull--back on the worldvolume of the brane, the combination 

\bea
C_4-C_2 \wedge B&=& e^{-\phi_0} \frac{l^4}{R}\sin^2{\alpha_1}\sin{\alpha_2}dt\wedge d\alpha_1 \wedge d\alpha_2 \wedge d\alpha_3+\non\\
&+& 4 R^4 e^{-\phi_0} \,\omega_1\wedge d\varphi_1 \wedge d\varphi_2 \wedge d\varphi_3
\eea

\noindent
is exactly the same as the R--R 4--form in the undeformed $AdS_5\times S^5$ space--time (recovered after setting the deformation parameters $\hat{\gamma}_i$ to zero). So, the independence of the deformation seems to be a feature of the Wess--Zumino term for a D3--brane configuration with vanishing worldvolume gauge field strength $F$ in this particular background\footnote{The authors of \cite{KISS} get a dependence on the deformation parameters but their conventions do not coincide with ours and with those of \cite{PolS,LM,F}.}. 

Can we speculate more on our $\hat{\gamma}_i$--independent results? Remember that we have pointed out that the existence of degenerate point--like and giant graviton states is not a new feature even in non--supersymmetric backgrounds \cite{PS} and in theories characterized by $B\neq 0$ \cite{CRK}. Moreover, our giant graviton solutions are classically BPS states in the deformed model, i.e. states that have the minimal energy for the given charge. The authors of \cite{FRT} discuss geodesics on $\hat{\gamma}_i$--deformed $\tilde{S}^5$ labeled by three conserved angular momenta $(J_1,J_2,J_3)$. These geodesics depend in general on the deformation parameters. In the standard $AdS_5\times S^5$ background all geodesics represent BPS states with energy $E$ equal to the total angular momentum $J=J_1+J_2+J_3$, while in the deformed case only few of them are characterized by this property. In particular, in the $\hat{\gamma}_i$--deformed model special solutions with energies that do not depend on the deformation parameters exist, i.e. they are the same as in the undeformed theory. This is the case for states labeled by $(J,0,0)$. We want to stress that our giant gravitons are $(J,0,0)$ BPS states and follow a geodesics of $\tilde{S}^5$, so that their classical independence on the deformation parameters is not a new feature. Moreover, studying giant gravitons on a deformed $(J,0,0)$ PP--wave, the authors of \cite{HM} also found a classical configuration independent of the deformation and with a spectrum of small fluctuations almost identical to the one obtained in section \ref{ggf}. This similar behavior could be an interesting point to study in detail.

The background we have studied breaks all the supersymmetries of $AdS_5\times S^5$ and so it should be dual to a non--supersymmetric but marginal deformation of the $\mathcal{N}=4$ $SU(N)$ SYM \cite{F}. More precisely, the gauge theory is conformal in the large $N$ limit \cite{FRT,K,DGK}, which we assume from now on. The bosonic part of the deformed YM has the following form\footnote{We use the notations of \cite{DGK}.}

\beq\label{V}
\mathcal{W}=\mbox{Tr}\left(-\frac{1}{2} \left[\Phi_i\,,\,\Phi_j\right]_{C_{ij}}\left[\bar{\Phi}^i\,,\,\bar{\Phi}^j\right]_{C_{ij}}+\frac{1}{4}\left[\Phi_i\,,\,\bar{\Phi}^i\right]\left[\Phi_j\,,\,\bar{\Phi}^j\right]\right)
\eeq

\noindent
where $\Phi_i$ are the three holomorphic scalars of $\mathcal{N}=4$ SYM. The deformation manifests itself in the modified commutators

\beq
\left[\Phi_i\,,\,\Phi_j\right]_{C_{ij}}\equiv e^{i C_{ij}}\Phi_i\,\Phi_j-e^{-i C_{ij}}\Phi_j\,\Phi_i\,,\qquad\qquad i,j=1,2,3
\eeq

\noindent
and similarly for the conjugate fields $\bar{\Phi}_i$. The matrix $C$ reads \cite{BR}

\beq\label{matrixc}
C=\pi
\left(
\begin{array}{ccc}
0 & -\gamma_3 & \gamma_2\\
\gamma_3 & 0 & -\gamma_1\\
-\gamma_2 & \gamma_1 & 0
\end{array}
\right)
\eeq

\noindent
The real deformation parameters $\hat{\gamma}_i$ appearing in (\ref{sdef}) are related to the $\gamma_i$ deformations on the gauge theory side (\ref{matrixc}) via the simple rescaling $\hat{\gamma}_i=R^2 \gamma_i$. The potential can be also obtained from the undeformed one by replacing the usual product $\Phi_i \Phi_j$ by the associative $\star$-product of \cite{LM,FRT}. 

The fact that the energy is independent of the deformation parameters is general and persists both in the case of unequal $\hat{\gamma}_i$ and in the $\mathcal{N}=1$ supersymmetric $\hat{\gamma}_i=\hat{\gamma}$ theory. In order to simplify our analysis of the dual $CFT$ picture of the giant gravitons, we restrict to the more studied $\mathcal{N}=1$ case where we are protected by supersymmetry. We have not checked that in the supersymmetric case our giant gravitons preserve some of the supersymmetries but the fact that they saturate a BPS bound is an indication of this feature. It would be interesting to prove this expectation. From now on we set $\gamma_i=\gamma$.   

Via $AdS/CFT$, states in supergravity are expected to map onto states of Yang--Mills theory on $\mathbb{R}\times S^3$ and the energy in space--time maps to energy in the field theory. Using the state--operator correspondence, the energy of states on $\mathbb{R}\times S^3$ maps to the dimension $\Delta=R E$ of operators on $\mathbb{R}^4$. In the undeformed case, the operators corresponding to (dual) giant gravitons have been first introduced in \cite{BBNS,CJR}. Our giant graviton solutions correspond to the case where we have only one non--vanishing angular momentum (a $(J,0,0)$ BPS state in the language of \cite{FRT}) and we should construct the dual operators on the $CFT$ side with only one holomorphic scalar field. Let $Z\equiv\Phi_1=\phi^5+i\phi^6$ be a complex combination of two of the six adjoint scalars in the YM theory, then in the undeformed case giant gravitons are dual to states created by a family of subdeterminants \cite{BBNS}

\beq\label{cop}
O_{Giant}=\frac{1}{J!}\epsilon_{i_1\,i_2\cdots i_J\,a_1\,a_2\cdots a_{N-J}}\,\epsilon^{l_1\,l_2\cdots l_J\,a_1\,a_2\cdots a_{N-J}}\,Z^{i_1}_{l_1}\,Z^{i_2}_{l_2}\cdots Z^{i_J}_{l_J}
\eeq

\noindent
Moreover, 

\beq\label{cops}
\tilde{O}_J=\frac{1}{J!}\sum_{\sigma \in \mathcal{S}_J}\,Z^{i_1}_{i_{\sigma(1)}}\,Z^{i_2}_{i_{\sigma(2)}}\cdots Z^{i_J}_{i_{\sigma(J)}}
\eeq

\noindent
with $\mathcal{S}_J$ the permutation group of length $J$, is supposed to describe a dual giant graviton in the undeformed theory \cite{CJR}. Once the deformation is turned on we are instructed to use the $\star$-product among the fields, so introducing a set of relative phases \cite{Pha}. However, the field content of the operators (\ref{cop}) and (\ref{cops}) implies a vanishing phase factor, and so we guess that the same operators could describe giant graviton states even in the $\gamma$--deformed theory \cite{HM}. All these operators form a good basis in the large $J\sim N$ limit and have classical scaling dimension $\Delta=J$, matching the results of sections \ref{confs} and \ref{confads}. This is an agreement between a strong and a weak coupling limits and so the operators (\ref{cop}) and (\ref{cops}) seem to be protected even in this less--supersymmetric case. Remember that single trace operators of the form $(J,0,0)$ are BPS states of the $\gamma$--deformed gauge theory which have zero anomalous dimension \cite{LM,NP} but we expect this property to hold also for the more complicated operators (\ref{cop}) and (\ref{cops}) because they can always be written as (Schur) polynomials in $Z$ \cite{CJR,CS2}\footnote{The authors of \cite{FRT} have shown that also in the non--supersymmetric case of three unequal $\gamma_i$, operators of the class $(J,0,0)$ are protected in the limit of large $N$. It is possible that the operators (\ref{cop}) and (\ref{cops}) could represent giant graviton states even in the non--supersymmetric case.}.

\subsection{Comments on the dual gauge theory picture of giant gravitons}

We have seen that the deformation seems to manifest itself in the vibration modes around the stable configurations. It would be very interesting to find the $CFT$ dual of these scalar fluctuations, as in \cite{BHLN,BBFH}. In general, most fluctuations of giant gravitons are not BPS and so from the field theory side we expect anomalous dimensions to develop quantum mechanically: The calculation would involve the full potential (\ref{V}) and of course the deformation parameters. From the brane side we read $\Delta=R E_{\omega}$, where $E_{\omega}$ is the excited energy of the giant graviton, i.e. if we switch to the quantum--mechanical system $E_{\omega}\sim E+\omega$ (with $\hbar=1$), and $\omega$ is a general fluctuation frequency. To be more explicit\footnote{The following analysis can be extended in the same way to the other giant fluctuations.} let us focus on the spectrum of small $AdS$ fluctuations when the giant graviton expands into the deformed 5--sphere $\tilde{S}^5$. The frequencies of the four modes are given by (\ref{omega}) 

\beq
\omega_k=\frac{\sqrt{(s+1)^2+\hat{\Gamma}^2}}{R}
\eeq

\noindent
with $Q_s^2=s(s+2)$. The radius $r_0$ of the spherical D3--brane enters in the definition of $\hat{\Gamma}^2$ (\ref{gammaq}) and the energy now reads

\beq\label{ene}
E_{\omega_k}=\frac{J}{R}+\frac{\sqrt{(s+1)^2+\lambda\,\Gamma^2}}{R}
\eeq

\noindent
We have used $R^4=4\pi e^{\phi_0}N=g_{YM}^2 N=\lambda$ and $\hat{\gamma}=R^2\gamma$, so that from (\ref{gammaq}) and $r_0^2=J R^2/N$, the relation 

\beq
\hat{\Gamma}^2=\lambda\,\Gamma^2=\lambda\,\left(1-\frac{J}{N}\right)\gamma^2(m_2- m_3)^2 
\eeq

\noindent
naturally follows. Note that for a maximal giant graviton $J=N$, $\Gamma^2=0$ and we recover the frequency obtained in the standard $AdS_5\times S^5$ case \cite{DJM}. If we want to find the dual description of these fluctuations, we can introduce suitable impurities in (\ref{cop}) as first proposed in \cite{BHLN} (the $\star$-product is implicit)

\beq\label{op}
\mathcal{O}^s_k \sim \epsilon_{i_1\,i_2\cdots i_J\,a_1\,a_2\cdots a_{N-J}}\,\epsilon^{l_1\,l_2\cdots l_J\,a_1\,a_2\cdots a_{N-J}}\,Z^{i_1}_{l_1}\,Z^{i_2}_{l_2}\cdots Z^{i_{J-1}}_{l_{J-1}}(W^s_k)^{i_J}_{l_J}
\eeq

\noindent
Here $W_k^s$ is a word built out of the $s$th symmetric traceless product of the other four scalars $\phi_i$ of the YM theory ($i=1,\cdots,4$) to match the scalar spherical harmonics of $S^3$ on the brane side. In order to consider fluctuations along the $AdS$ directions we have to include a covariant derivative $D_k$ in the word, so the index $k=1 \cdots 4$ refers to the four Cartesian directions of $\mathbb{R}^4$ in radial quantization of $\mathbb{R}\times S^3$. We stress that the deformation parameters introduce a dependence on the 't Hooft coupling $\lambda$ and, if the $AdS/CFT$ correspondence holds, the energy $E_{\omega_k}$ gives the scaling dimension of $\mathcal{O}^s_k$ in the limit of large 't Hooft coupling

\beq\label{dimtot}
\Delta=J+\sqrt{(s+1)^2+\lambda\,\Gamma^2}
\eeq

\noindent
We do not exclude the possibility that the interactions of the Yang--Mills theory do produce a perturbative (weak coupling constant $\lambda\ll 1$) anomalous dimension for the operators just introduced, related to that predicted by the other side of the correspondence. This is a heuristic discussion, since the precise form of a general operator of the type (\ref{op}) is still unknown. Moreover, we are now talking about non--protected quantities and a direct comparison is a very difficult task because we are facing a strong/weak coupling duality. If we want to match the results, it is simpler to study the correspondence in novel limits, for example where quantum numbers become large with $N$ \cite{BMN}.

\subsection{Dual giants and semi--classical solutions of $CFT$}

The fluctuations around dual giants can be similarly described using operators on the field theory side (see the recent \cite{CSd}). However, a more efficient approach is to identify a classical field theory configuration which encodes the same properties of the spherical brane in $AdS$ \cite{HHI} and then try to study fluctuations around this solution similarly to \cite{BBFH}. The idea is to work with the bosonic part of the dual $CFT$ which lives on the boundary of $AdS_5$, namely on $\mathbb{R}\times S^3$ with metric $ds^2=h_{\mu\nu} dx^\mu dx^\nu$

\beq
S=-\frac{1}{g^2_{YM}}\int d^4x \sqrt{-h}\,\left[\mbox{Tr}\left(\partial^\mu \bar{\Phi}_i\partial_\mu \Phi^i+\frac{1}{R^2}\bar{\Phi}_i\Phi^i\right)+\mathcal{W}\,\right]
\eeq

\noindent
where $\mathcal{W}$ is defined in (\ref{V}) with $\gamma_i=\gamma$. Since the background is of the form $\mathbb{R}\times S^3$, the conformal invariance of the theory imposes a mass term for $\Phi_i$ and $R$ is the radius of $AdS_5$. By rescaling the $\Phi_i$ fields

\beq
\Phi_i(t,\Omega)\,\rightarrow\,\sqrt{\frac{g^2_{YM}N}{4 \pi^2 R^2}}\,\Phi_i(t,\Omega)
\eeq 

\noindent
the action can be rewritten as

\beq\label{act}
S=\frac{N}{4 \pi^2 R^2}\int d^4x \sqrt{-h}\,\left[\mbox{Tr}\left(-\partial^\mu\bar{\Phi}_i\partial_\mu{\Phi}^i-\frac{1}{R^2}\bar{\Phi}_i\Phi^i\right)+\mathcal{W}_\lambda\,\right]
\eeq

\noindent
The rescaled potential $\mathcal{W}_\lambda$ is

\beq\label{VG}
\mathcal{W}_\lambda=\mbox{Tr}\left(\frac{\lambda}{8 \pi^2 R^2} \left[\Phi_i\,,\,\Phi_j\right]_{C_{ij}}\left[\bar{\Phi}^i\,,\,\bar{\Phi}^j\right]_{C_{ij}}-\frac{\lambda}{16 \pi^2 R^2}\left[\Phi_i\,,\,\bar{\Phi}^i\right]\left[\Phi_j\,,\,\bar{\Phi}^j\right]\right)
\eeq

\noindent
The matrix $C_{ij}$ is defined in (\ref{matrixc}) with now $\gamma_i=\gamma$. Next, we consider the ansatz 

\beq\label{cla}
\Phi_1(t,\Omega)=\mbox{diag}(\eta,0,0,\cdots,0)e^{i\theta(t)}\quad\mbox{with}\quad \eta=\mbox{const.}\,,\qquad\qquad \Phi_{2,3}(t,\Omega)=0
\eeq

\noindent
which is well defined in the large $N$ limit of the $SU(N)$ theory \cite{HHI}. The Lagrangian turns out to be

\beq
L=\frac{N R}{2}\left(\eta^2\dot{\theta}^2-\frac{\eta^2}{R^2}\right)
\eeq 

\noindent
We see that the angular momentum $J=\partial L/\partial \dot{\theta}$ is conserved and the energy

\beq
E=J \dot{\theta}-L=\frac{J^2}{2 N R \eta^2}+\frac{N \eta^2}{2 R}
\eeq

\noindent
is minimized at $\eta^2=\eta_0^2=J/N$ where its value is $E=J/R$. So, we have found a classical configuration in this truncated $CFT$ which has the same properties of the spherical brane in $AdS_5$.

The transverse fluctuations of dual giants are represented in the gauge theory by modes of the scalars $\phi_i$ for $i=1,\cdots,4$ as explained in the previous section. The coordinates $(\rho_i,\varphi_i)$ which parametrize the deformed sphere $\tilde{S}^5$ (\ref{sdef}) correspond to the three complex scalars $\Phi_i$ of Yang--Mills theory and in particular the dictionary tells us that $\Phi_i=\rho_i e^{i\varphi_i}$. On the supergravity side the modified ansatz (\ref{ans}) yields to $\rho_{2,3}\sim \varepsilon \delta r/\sqrt{2}$ and so if we want to translate this vibrations in the dual $CFT$ it seems natural to consider diagonal fluctuations of the form

\beq
\Phi_2(t,\Omega)=\Phi_3(t,\Omega)=\varepsilon\,\mbox{diag}(\frac{\delta\rho(t,\Omega)}{\sqrt{2}},0,0,\cdots,0)
\eeq 

\noindent
Moreover, we note that in this $CFT$ analysis $\eta$ covers the role of the radius of the giant, while $\dot{\theta}$ is the angular velocity; in fact at the minimum of the energy its value is $\dot{\theta}=\theta_0=1/R$ as in (\ref{omega0a}). So we guess that the study of small fluctuations in radius and in the orientation of angular momentum could be performed thanks to the modified ansatz

\beq
\Phi_1(t,\Omega)=\mbox{diag}(\eta+\varepsilon\,\delta\eta(t,\Omega),0,0,\cdots,0)e^{i(\theta(t)+\varepsilon\,\delta\theta(t,\Omega))}
\eeq

\noindent
Exactly as in section \ref{vib}, we study the action up to second order in $\varepsilon$ and we expand the generic perturbation $\delta x(t,\Omega)$ in spherical harmonics

\beq
\delta x(t,\Omega)=A_x e^{-i \tilde{\omega}_x t} \mathcal{Y}_s(\Omega)
\eeq

\noindent
The calculation runs parallel to that of section \ref{dggf} so we are free to omit the details; we only stress that the linear term in $\varepsilon$ vanishes when evaluated in the classical vacuum and the first commutator in (\ref{VG}) covers a crucial role in what follows. The $\delta \eta$, $\delta \theta$ perturbations are coupled and the resulting frequencies are

\bea
\tilde{\omega}_{\pm}^2&=&\frac{1}{R^2}\left(2+Q_s^2\pm 2\sqrt{1+Q_s^2}\right)
\eea

\noindent
in perfect agreement with (\ref{omegapma}). The $\delta \rho$ perturbation decouples from $\delta \eta$, $\delta \theta$ and has a frequency

\beq\label{omegaro}
\tilde{\omega}_\rho^2=\frac{1}{R^2}\left(1+Q^2_s+\frac{\lambda}{4 \pi^2}|q-\bar{q}|^2\frac{J}{N}\right)
\eeq

\noindent
where we have defined $q=e^{i\pi\gamma}$. Note that the frequency (\ref{omegaro}) is very similar to the exact anomalous dimension obtained in \cite{MPSZ}. When the deformation parameter is set to zero ($q=\bar{q}=1$), we recover the frequencies obtained in the undeformed theory \cite{BBFH}. Because of the $\lambda$--dependence of (\ref{omegaro}) we have to be careful in comparing it to the result of section \ref{dggf}. Quantum mechanically, the energy (in units of $1/R$ and with $\hbar=1$) has the form

\beq\label{EC}
E_{CFT}=J+\sqrt{1+Q^2_s+\frac{\lambda}{4 \pi^2}|q-\bar{q}|^2\frac{J}{N}}
\eeq 

\noindent
On the other hand, from the value of the small fluctuation frequency given in (\ref{omegar}) and with $\hat{\gamma}_2^2=\hat{\gamma}_3^2=\hat{\gamma}^2=\lambda\gamma^2$, the energy of the brane is

\beq\label{ES}
E_{BRANE}=J+\sqrt{1+Q^2_s+\lambda\left(1+\frac{J}{N}\right)\gamma^2}
\eeq 

\noindent
What happened? The two energies are remarkably similar but of course we have to check the regime of validity of our analysis of the small vibrations, both in the gauge theory and in the supergravity side. To be more precise, the energy $E_{BRANE}$ (\ref{ES}) is a well defined quantity at large $\lambda$ and in the small $\gamma$ limit, with $\hat{\gamma}^2=\lambda\gamma^2$ fixed \cite{LM}. The $CFT$ energy (\ref{EC}) was computed for small $\lambda$, where the semi--classical description of the Yang--Mills theory becomes reliable, and at arbitrary $q$. So we expect a function to exist which smoothly interpolates between the weak coupling result (\ref{EC}) and the strong coupling one (\ref{ES}). Note that if we expand the $|q-\bar{q}|^2$ term into the square root of (\ref{EC}) for a particularly small value of $\gamma$, we obtain 

\beq
E_{CFT}\sim J+\sqrt{1+Q^2_s+\lambda\, \gamma^2\,\frac{J}{N}}
\eeq 

\noindent 
On the other hand, if $J/N\gg 1$ we can safely ignore the 1 appearing in (\ref{ES}) and up to their regime of validity, the two energies are identical. This is the same limit studied in \cite{HHI} following the work of \cite{SW}, to show that for large values of $J/N$ the leading term of the Dirac--Born--Infeld and Wess--Zumino action of a brane in $AdS_5$ exactly matches the $CFT$ action. We leave the complete understanding of these features for future works. Another useful strategy could be the one used in \cite{Bq}. 

Our $CFT$ analysis applies equally well to the case of unequal $\gamma_i$ and reproduces the $\gamma_{2,3}$ behavior obtained on the brane side. So, let us conclude noting that in particular the authors of \cite{FRT} and \cite{DGK} have found non--trivial examples where implications of the $AdS/CFT$ duality are observed even in the non--supersymmetric case and where the non--renormalization theorem seems not to be dictated by supersymmetry. We do not exclude a possible extension of this $AdS/CFT$ comparison to the more general case of unequal $\gamma_i$ deformation parameters.

\section{Conclusions}

In this paper we have considered giant graviton configurations on the Type IIB supergravity background which can be obtained by a non--supersymmetric but marginal three--parameter deformation of the original $AdS_5\times S^5$ solution. In particular, we have shown the existence of giants which are energetically indistinguishable from the point graviton, even in absence of supersymmetry. This feature holds for both the two sets of giant graviton solutions, namely when the D3--brane expands into the deformed 5--sphere part of the geometry and when it blows up into $AdS_5$. The (dual) giant dynamics turns out to be independent of the deformation parameters with a behavior which is exactly the same found in the undeformed theory. The deformation of the background affects both the NS--NS and the R--R sectors. The D3--brane couples to the two and four forms but with a precise mechanism which exactly compensates the changes induced by the deformation. More striking, this complete cancellation of the deformation parameters does not depend on their values and remains valid in the presence of unequal $\hat{\gamma}_i$ (the non--supersymmetric case) and in the special case $\hat{\gamma}_i=\hat{\gamma}$, corresponding to the supersymmetric Lunin--Maldacena deformation. In order to understand the stability of the configurations we have found, we have also performed a systematic study of the spectrum of small fluctuations around the giant graviton solutions. This is where the deformation manifests itself providing the first important difference with respect to the undeformed case. In fact, the deformed spectrum turns out to depend on the radius of the (dual) giant which is always coupled to the deformation parameters. Despite this fact, the deformation enters into the spectrum as a positive contribution and the frequencies do not allow tachyonic modes. The (dual) giant gravitons are perturbatively stable and this characteristic works in favor of the perfect quantitative agreement between the gauge theory and the string theory found in \cite{KISS}. Finally, restricting to the supersymmetric case of equal $\hat{\gamma}_i$, we have proposed qualitative and quantitative comparisons obtained from the dual gauge theory picture, generalizing what is known in the original undeformed correspondence. In the case of dual giant gravitons, a semi--classical $CFT$ picture seems to capture a lot of the physics of the brane configuration, giving the correct energy, angular momentum and a remarkable similar spectrum of small fluctuations.    

The study of giant graviton dynamics is certain a fascinating subject. One of their most striking features is their ability to relate UV and IR regimes by enlarging their size with the increasing of the energy. Another interesting feature of giant graviton solutions is  their stability even in a non--supersymmetric background. Further investigations of this property could give new insight in the understanding of the role played by supersymmetry in the gauge/gravity dualities.

%\vspace{1.0cm}

\section*{Acknowledgments}

\noindent 
I would like to thank in primis S.~Penati, and A.~Butti, M.~Caldarelli, R.~de~Mello~Koch, D.~Forcella, O.~Lunin, A.~Mariotti, L.~Mazzanti, A.~Romagnoni, A.~Sartirana, M.~Smolic, G.~Tartaglino--Mazzucchelli and A.~Zaffaroni for useful discussions.

This work has been supported in part by INFN, PRIN prot. $2005-024045-004$ and the European Commission RTN program MRTN--CT--2004--005104.

\newpage
%%%%%%%%%%%%--References--%%%%%%%%%%%%%%%%%%%%%


\begin{thebibliography}{99}
\parskip-2pt

\bibitem{AdS/CFT}
J.~M.~Maldacena,
%``The large N limit of superconformal field theories and supergravity,''
Adv.\ Theor.\ Math.\ Phys.\  {\bf 2} (1998) 231 [Int.\ J.\ Theor.\
Phys.\  {\bf 38} (1999) 1113] [arXiv:hep-th/9711200].\\
%%CITATION = HEP-TH 9711200;%%
S.~S.~Gubser, I.~R.~Klebanov and A.~M.~Polyakov,
%``Gauge theory correlators from non-critical string theory,''
Phys.\ Lett.\ B {\bf 428} (1998) 105 [arXiv:hep-th/9802109].\\
%%CITATION = HEP-TH 9802109;%%
E.~Witten,
%``Anti-de Sitter space and holography,''
Adv.\ Theor.\ Math.\ Phys.\  {\bf 2} (1998) 253
[arXiv:hep-th/9802150].
%%CITATION = HEP-TH 9802150;%%


\bibitem{LS}
R.~G.~Leigh and M.~J.~Strassler,
%``Exactly marginal operators and duality in four-dimensional N=1 supersymmetric
%gauge theory,''
Nucl.\ Phys.\ B {\bf 447} (1995) 95 [arXiv:hep-th/9503121].
%%CITATION = HEP-TH 9503121;%%


\bibitem{LM}
O.~Lunin and J.~Maldacena,
%``Deforming field theories with U(1) x U(1) global symmetry and their gravity
%duals,''
JHEP {\bf 0505} (2005) 033 [arXiv:hep-th/0502086].
%%CITATION = HEP-TH 0502086;%%


\bibitem{F}
S.~Frolov,
%``Lax pair for strings in Lunin-Maldacena background,''
JHEP {\bf 0505}, 069 (2005)
[arXiv:hep-th/0503201].
%%CITATION = HEP-TH 0503201;%%


\bibitem{RVY}
R.~C.~Rashkov, K.~S.~Viswanathan and Y.~Yang,
%``Generalization Of The Lunin-Maldacena Transformation On The Ads(5) X S**5
%Background,''
Phys.\ Rev.\ D {\bf 72}, 106008 (2005)
[arXiv:hep-th/0509058].
%%CITATION = HEP-TH 0509058;%%


\bibitem{AAF}
L.~F.~Alday, G.~Arutyunov and S.~Frolov,
%``Green-Schwarz strings in TsT-transformed backgrounds,''
JHEP {\bf 0606}, 018 (2006)
[arXiv:hep-th/0512253].
%%CITATION = HEP-TH 0512253;%%


\bibitem{T}
S.~A.~Frolov, R.~Roiban and A.~A.~Tseytlin,
%``Gauge - string duality for superconformal deformations of N = 4 super
%Yang-Mills theory,''
JHEP {\bf 0507}, 045 (2005)
[arXiv:hep-th/0503192].
%%CITATION = HEP-TH 0503192;%%


\bibitem{My}
R.~C.~Myers,
%``Dielectric-branes,''
JHEP {\bf 9912}, 022 (1999)
[arXiv:hep-th/9910053].
%%CITATION = HEP-TH 9910053;%%


\bibitem{MGST}
J.~McGreevy, L.~Susskind and N.~Toumbas,
%``Invasion Of The Giant Gravitons From Anti-De Sitter Space,''
JHEP {\bf 0006}, 008 (2000)
[arXiv:hep-th/0003075].
%%CITATION = HEP-TH 0003075;%%


\bibitem{GMT}
M.~T.~Grisaru, R.~C.~Myers and O.~Tafjord,
%``SUSY and Goliath,''
JHEP {\bf 0008}, 040 (2000)
[arXiv:hep-th/0008015].
%%CITATION = HEP-TH 0008015;%%


\bibitem{HHI}
A.~Hashimoto, S.~Hirano and N.~Itzhaki,
%``Large branes in AdS and their field theory dual,''
JHEP {\bf 0008}, 051 (2000)
[arXiv:hep-th/0008016].
%%CITATION = HEP-TH 0008016;%%


\bibitem{BBNS}
V.~Balasubramanian, M.~Berkooz, A.~Naqvi and M.~J.~Strassler,
%``Giant gravitons in conformal field theory,''
JHEP {\bf 0204}, 034 (2002)
[arXiv:hep-th/0107119].
%%CITATION = HEP-TH 0107119;%%


\bibitem{CJR}
S.~Corley, A.~Jevicki and S.~Ramgoolam,
%``Exact correlators of giant gravitons from dual N = 4 SYM theory,''
Adv.\ Theor.\ Math.\ Phys.\  {\bf 5}, 809 (2002)
[arXiv:hep-th/0111222].
%%CITATION = HEP-TH 0111222;%%


\bibitem{others}
%\cite{Aharony:2002nd}
%\bibitem{Aharony:2002nd}
O.~Aharony, Y.~E.~Antebi, M.~Berkooz and R.~Fishman,
%``'Holey sheets': Pfaffians and subdeterminants as D-brane operators in large
%N gauge theories,''
JHEP {\bf 0212}, 069 (2002)
[arXiv:hep-th/0211152].\\
%%CITATION = HEP-TH 0211152;%%
%\cite{Berenstein:2003ah}
%\bibitem{Berenstein:2003ah}
D.~Berenstein,
%``Shape and holography: Studies of dual operators to giant gravitons,''
Nucl.\ Phys.\ B {\bf 675}, 179 (2003)
[arXiv:hep-th/0306090].\\
%%CITATION = HEP-TH 0306090;%%
%\cite{deMelloKoch:2004ws}
%\bibitem{deMelloKoch:2004ws}
R.~de Mello Koch and R.~Gwyn,
%``Giant graviton correlators from dual SU(N) super Yang-Mills theory,''
JHEP {\bf 0411}, 081 (2004)
[arXiv:hep-th/0410236].\\
%%CITATION = HEP-TH 0410236;%%
%\cite{Berenstein:2005vf}
%\bibitem{Berenstein:2005vf}
D.~Berenstein and S.~E.~Vazquez,
%``Integrable open spin chains from giant gravitons,''
JHEP {\bf 0506}, 059 (2005)
[arXiv:hep-th/0501078].\\
%%CITATION = HEP-TH 0501078;%%
%\cite{Berenstein:2005fa}
%\bibitem{Berenstein:2005fa}
D.~Berenstein, D.~H.~Correa and S.~E.~Vazquez,
%``Quantizing open spin chains with variable length: An example from giant
%gravitons,''
Phys.\ Rev.\ Lett.\  {\bf 95}, 191601 (2005)
[arXiv:hep-th/0502172].\\
%%CITATION = HEP-TH 0502172;%%
%\cite{Agarwal:2006gc}
%\bibitem{Agarwal:2006gc}
A.~Agarwal,
%``Open spin chains in super Yang-Mills at higher loops: Some potential
%problems with integrability,''
JHEP {\bf 0608}, 027 (2006)
[arXiv:hep-th/0603067].\\
%%CITATION = HEP-TH 0603067;%%
%\cite{Berenstein:2006qk}
%\bibitem{Berenstein:2006qk}
D.~Berenstein, D.~H.~Correa and S.~E.~Vazquez,
%``A study of open strings ending on giant gravitons, spin chains and
%integrability,''
JHEP {\bf 0609}, 065 (2006)
[arXiv:hep-th/0604123].
%%CITATION = HEP-TH 0604123;%%


\bibitem{Bere}
D.~Berenstein,
%``A toy model for the AdS/CFT correspondence,''
JHEP {\bf 0407}, 018 (2004)
[arXiv:hep-th/0403110].
%%CITATION = HEP-TH 0403110;%%


\bibitem{LLM}
H.~Lin, O.~Lunin and J.~M.~Maldacena,
%``Bubbling AdS space and 1/2 BPS geometries,''
JHEP {\bf 0410}, 025 (2004)
[arXiv:hep-th/0409174].
%%CITATION = HEP-TH 0409174;%%


\bibitem{general}
%\cite{Das:2000fu}
%\bibitem{Das:2000fu}
S.~R.~Das, A.~Jevicki and S.~D.~Mathur,
%``Giant gravitons, BPS bounds and noncommutativity,''
Phys.\ Rev.\ D {\bf 63}, 044001 (2001)
[arXiv:hep-th/0008088].\\
%%CITATION = HEP-TH 0008088;%%
%\cite{Lee:2000vd}
%\bibitem{Lee:2000vd}
J.~Lee,
%``Tunneling between the giant gravitons in AdS(5) x S(5),''
Phys.\ Rev.\ D {\bf 64}, 046012 (2001)
[arXiv:hep-th/0010191].\\
%%CITATION = HEP-TH 0010191;%%
%\cite{Sadri:2003mx}
%\bibitem{Sadri:2003mx}
D.~Sadri and M.~M.~Sheikh-Jabbari,
%``Giant hedge-hogs: Spikes on giant gravitons,''
Nucl.\ Phys.\ B {\bf 687}, 161 (2004)
[arXiv:hep-th/0312155].\\
%%CITATION = HEP-TH 0312155;%%
%\cite{Arapoglu:2003ti}
%\bibitem{Arapoglu:2003ti}
S.~Arapoglu, N.~S.~Deger, A.~Kaya, E.~Sezgin and P.~Sundell,
%``Multi-spin giants,''
Phys.\ Rev.\ D {\bf 69}, 106006 (2004)
[arXiv:hep-th/0312191].\\
%%CITATION = HEP-TH 0312191;%%
%\bibitem{CS}
M.~M.~Caldarelli and P.~J.~Silva,
%``Multi giant graviton systems, SUSY breaking and CFT,''
JHEP {\bf 0402}, 052 (2004)
[arXiv:hep-th/0401213].\\
%%CITATION = HEP-TH 0401213;%%
%\cite{Prokushkin:2004pv}
%\bibitem{Prokushkin:2004pv}
S.~Prokushkin and M.~M.~Sheikh-Jabbari,
%``Squashed giants: Bound states of giant gravitons,''
JHEP {\bf 0407}, 077 (2004)
[arXiv:hep-th/0406053].\\
%%CITATION = HEP-TH 0406053;%%
%\cite{Janssen:2004cd}
%\bibitem{Janssen:2004cd}
B.~Janssen, Y.~Lozano and D.~Rodriguez-Gomez,
%``Giant gravitons and fuzzy CP(2),''
Nucl.\ Phys.\ B {\bf 712}, 371 (2005)
[arXiv:hep-th/0411181].\\
%%CITATION = HEP-TH 0411181;%%
%\cite{Janssen:2004gd}
%\bibitem{Janssen:2004gd}
B.~Janssen, Y.~Lozano and D.~Rodriguez-Gomez,
%``Giant gravitons as fuzzy manifolds,''
arXiv:hep-th/0412037.\\
%%CITATION = HEP-TH 0412037;%%
%\cite{Huang:2006ku}
%\bibitem{Huang:2006ku}
W.~H.~Huang,
%``Electric / magnetic field deformed giant gravitons in Melvin geometry,''
Phys.\ Lett.\ B {\bf 635}, 141 (2006)
[arXiv:hep-th/0602019].\\
%%CITATION = HEP-TH 0602019;%%
%\cite{Biswas:2006tj}
%\bibitem{Biswas:2006tj}
I.~Biswas, D.~Gaiotto, S.~Lahiri and S.~Minwalla,
%``Supersymmetric states of N = 4 Yang-Mills from giant gravitons,''
arXiv:hep-th/0606087.\\
%%CITATION = HEP-TH 0606087;%%
%\cite{Mandal:2006tk}
%\bibitem{Mandal:2006tk}
G.~Mandal and N.~V.~Suryanarayana,
%``Counting 1/8-BPS dual-giants,''
arXiv:hep-th/0606088.\\
%%CITATION = HEP-TH 0606088;%%
%\cite{Martelli:2006vh}
%\bibitem{Martelli:2006vh}
D.~Martelli and J.~Sparks,
%``Dual giant gravitons in Sasaki-Einstein backgrounds,''
arXiv:hep-th/0608060.\\
%%CITATION = HEP-TH 0608060;%%
%\cite{Basu:2006id}
%\bibitem{Basu:2006id}
A.~Basu and G.~Mandal,
%``Dual giant gravitons in AdS(m) x Y**n (Sasaki-Einstein),''
arXiv:hep-th/0608093.
%%CITATION = HEP-TH 0608093;%%


\bibitem{KISS}
R.~de Mello Koch, N.~Ives, J.~Smolic and M.~Smolic,
%``Unstable giants,''
Phys.\ Rev.\ D {\bf 73}, 064007 (2006)
[arXiv:hep-th/0509007].
%%CITATION = HEP-TH 0509007;%%


\bibitem{HM}
A.~Hamilton and J.~Murugan,
%``Giant Gravitons on Deformed pp-waves,''
arXiv:hep-th/0609135.
%%CITATION = HEP-TH 0609135;%%


\bibitem{FRT}
S.~A.~Frolov, R.~Roiban and A.~A.~Tseytlin,
%``Gauge-string duality for (non)supersymmetric deformations of N = 4 super
%Yang-Mills theory,''
Nucl.\ Phys.\ B {\bf 731}, 1 (2005)
[arXiv:hep-th/0507021].
%%CITATION = HEP-TH 0507021;%%


\bibitem{Sz}
R.~J.~Szabo,
%``BUSSTEPP lectures on string theory: An introduction to string theory  and
%D-brane dynamics,''
arXiv:hep-th/0207142.
%%CITATION = HEP-TH 0207142;%%


\bibitem{PS}
D.~C.~Page and D.~J.~Smith,
%``Giant gravitons in non-supersymmetric backgrounds,''
JHEP {\bf 0207}, 028 (2002)
[arXiv:hep-th/0204209].
%%CITATION = HEP-TH 0204209;%%


\bibitem{CRK}
J.~M.~Camino and A.~V.~Ramallo,
%``Giant gravitons with NSNS B field,''
JHEP {\bf 0109} (2001) 012
[arXiv:hep-th/0107142].\\
%%CITATION = HEP-TH 0107142;%%
J.~Y.~Kim,
%``Stability of giant gravitons with NSNS B field,''
Phys.\ Lett.\ B {\bf 529}, 150 (2002)
[arXiv:hep-th/0109192].
%%CITATION = HEP-TH 0109192;%%


\bibitem{DJM}
S.~R.~Das, A.~Jevicki and S.~D.~Mathur,
%``Vibration modes of giant gravitons,''
Phys.\ Rev.\ D {\bf 63}, 024013 (2001)
[arXiv:hep-th/0009019].
%%CITATION = HEP-TH 0009019;%%


\bibitem{O}
P.~Ouyang,
%``Semiclassical quantization of giant gravitons,''
arXiv:hep-th/0212228.
%%CITATION = HEP-TH 0212228;%%


\bibitem{PolS}
J.~Polchinski and M.~J.~Strassler,
%``The string dual of a confining four-dimensional gauge theory,''
arXiv:hep-th/0003136.
%%CITATION = HEP-TH 0003136;%%


\bibitem{K}
V.~V.~Khoze,
%``Amplitudes in the beta-deformed conformal Yang-Mills,''
JHEP {\bf 0602}, 040 (2006)
[arXiv:hep-th/0512194].
%%CITATION = HEP-TH 0512194;%%


\bibitem{DGK}
C.~Durnford, G.~Georgiou and V.~V.~Khoze,
%``Instanton test of non-supersymmetric deformations of the AdS(5) x S**5,''
JHEP {\bf 0609}, 005 (2006)
[arXiv:hep-th/0606111].
%%CITATION = HEP-TH 0606111;%%


\bibitem{BR}
N.~Beisert and R.~Roiban,
%``Beauty and the twist: The Bethe ansatz for twisted N = 4 SYM,''
JHEP {\bf 0508}, 039 (2005)
[arXiv:hep-th/0505187].
%%CITATION = HEP-TH 0505187;%%


\bibitem{Pha}
T.~Mateos,
%``Marginal deformation of N = 4 SYM and Penrose limits with continuum
%spectrum,''
JHEP {\bf 0508}, 026 (2005)
[arXiv:hep-th/0505243].\\
%%CITATION = HEP-TH 0505243;%%
R.~de Mello Koch, J.~Murugan, J.~Smolic and M.~Smolic,
%``Deformed PP-waves from the Lunin-Maldacena background,''
JHEP {\bf 0508}, 072 (2005)
[arXiv:hep-th/0505227].\\
%%CITATION = HEP-TH 0505227;%%
D.~Bundzik,
%``Star product and the general Leigh-Strassler deformation,''
arXiv:hep-th/0608215.
%%CITATION = HEP-TH 0608215;%%


\bibitem{NP}
V.~Niarchos and N.~Prezas,
%``BMN operators for N = 1 superconformal Yang-Mills theories and associated
%string backgrounds,''
JHEP {\bf 0306}, 015 (2003)
[arXiv:hep-th/0212111].\\
%%CITATION = HEP-TH 0212111;%%
D.~Z.~Freedman and U.~Gursoy,
%``Comments on the beta-deformed N = 4 SYM theory,''
JHEP {\bf 0511}, 042 (2005)
[arXiv:hep-th/0506128].\\
%%CITATION = HEP-TH 0506128;%%
S.~Penati, A.~Santambrogio and D.~Zanon,
%``Two-point correlators in the beta-deformed N = 4 SYM at the next-to-leading
%order,''
JHEP {\bf 0510}, 023 (2005)
[arXiv:hep-th/0506150].\\
%%CITATION = HEP-TH 0506150;%%
%\cite{Rossi:2005mr}
%\bibitem{Rossi:2005mr}
G.~C.~Rossi, E.~Sokatchev and Y.~S.~Stanev,
%``New results in the deformed N = 4 SYM theory,''
Nucl.\ Phys.\ B {\bf 729}, 581 (2005)
[arXiv:hep-th/0507113].\\
%%CITATION = HEP-TH 0507113;%%
A.~Mauri, S.~Penati, M.~Pirrone, A.~Santambrogio, D.~Zanon,
%``On the perturbative chiral ring for marginally deformed N = 4 SYM
%theories,''
JHEP {\bf 0608}, 072 (2006)
[arXiv:hep-th/0605145].
%%CITATION = HEP-TH 0605145;%%


\bibitem{CS2}
M.~M.~Caldarelli and P.~J.~Silva,
%``Giant gravitons in AdS/CFT. I: Matrix model and back reaction,''
JHEP {\bf 0408}, 029 (2004)
[arXiv:hep-th/0406096].
%%CITATION = HEP-TH 0406096;%%


\bibitem{BHLN}
V.~Balasubramanian, M.~x.~Huang, T.~S.~Levi and A.~Naqvi,
%``Open strings from N = 4 super Yang-Mills,''
JHEP {\bf 0208}, 037 (2002)
[arXiv:hep-th/0204196].
%%CITATION = HEP-TH 0204196;%%


\bibitem{BBFH}
V.~Balasubramanian, D.~Berenstein, B.~Feng and M.~x.~Huang,
%``D-branes in Yang-Mills theory and emergent gauge symmetry,''
JHEP {\bf 0503} (2005) 006
[arXiv:hep-th/0411205].
%%CITATION = HEP-TH 0411205;%%


\bibitem{BMN}
D.~Berenstein, J.~M.~Maldacena and H.~Nastase,
%``Strings in flat space and pp waves from N = 4 super Yang Mills,''
JHEP {\bf 0204} (2002) 013
[arXiv:hep-th/0202021].\\
%%CITATION = HEP-TH 0202021;%%
S.~S.~Gubser, I.~R.~Klebanov and A.~M.~Polyakov,
%``A semi-classical limit of the gauge/string correspondence,''
Nucl.\ Phys.\ B {\bf 636}, 99 (2002)
[arXiv:hep-th/0204051].\\
%%CITATION = HEP-TH 0204051;%%
S.~Frolov and A.~A.~Tseytlin,
%``Semiclassical quantization of rotating superstring in AdS(5) x S(5),''
JHEP {\bf 0206}, 007 (2002)
[arXiv:hep-th/0204226].\\
%%CITATION = HEP-TH 0204226;%%
S.~Frolov and A.~A.~Tseytlin,
%``Multi-spin string solutions in AdS(5) x S**5,''
Nucl.\ Phys.\ B {\bf 668}, 77 (2003)
[arXiv:hep-th/0304255].\\
%%CITATION = HEP-TH 0304255;%%
S.~Frolov and A.~A.~Tseytlin,
%``Quantizing three-spin string solution in AdS(5) x S**5,''
JHEP {\bf 0307}, 016 (2003)
[arXiv:hep-th/0306130].
%%CITATION = HEP-TH 0306130;%%


\bibitem{CSd}
D.~H.~Correa and G.~A.~Silva,
%``Dilatation operator and the super Yang-Mills duals of open strings on AdS
%giant gravitons,''
arXiv:hep-th/0608128.
%%CITATION = HEP-TH 0608128;%%


\bibitem{MPSZ}
A.~Mauri, S.~Penati, A.~Santambrogio and D.~Zanon,
%``Exact results in planar N = 1 superconformal Yang-Mills theory,''
JHEP {\bf 0511}, 024 (2005)
[arXiv:hep-th/0507282].
%%CITATION = HEP-TH 0507282;%%


\bibitem{SW}
N.~Seiberg and E.~Witten,
%``The D1/D5 system and singular CFT,''
JHEP {\bf 9904}, 017 (1999)
[arXiv:hep-th/9903224].
%%CITATION = HEP-TH 9903224;%%


\bibitem{Bq}
D.~Berenstein and D.~H.~Correa,
%``Emergent geometry from q-deformations of N = 4 super Yang-Mills,''
JHEP {\bf 0608}, 006 (2006)
[arXiv:hep-th/0511104].
%%CITATION = HEP-TH 0511104;%%




\end{thebibliography}
\end{document}